# Low Loss and Magnetic Field-tuned Superconducting THz Metamaterial

**Biaobing Jin,[1,*] Caihong Zhang,[1] Sebastian Engelbrecht,[2] Andrei Pimenov,[2] Jingbo Wu,[1] Qinyin Xu,[1] Chunhai Cao,[1] Jian Chen,[1] Weiwei Xu,[1] Lin Kang,[1] and Peiheng Wu[1]**

[1] *Research Institute of Superconductor Electronics (RISE), School of Electronic Science and Engineering, Nanjing University, Nanjing 210093, China.*

[2] *Experimentelle Physik IV, Universität Würzburg, Am Hubland, D-97074 Würzburg, Germany*

*\*bbjin@nju.edu.cn*

**Abstract:** Superconducting terahertz (THz) metamaterial (MM) made from superconducting Nb film has been investigated using a continuous-wave THz spectroscopy with a superconducting split-coil magnet. The obtained quality factors of the resonant modes at 132 GHz and 450 GHz are about three times as large as those calculated for a metal THz MM operating at 1 K, which indicates that superconducting THz MM is a very nice candidate to achieve low loss performance. In addition, the magnetic field-tuning on superconducting THz MM is also demonstrated, which offer an alternative tuning method apart from the existed electric, optical and thermal tuning on THz MM.

©2010 Optical Society of America

**OCIS codes:** (160.3918) Metamaterials; (260.5740) Resonance; (300.6495) spectroscopy, terahertz

## 1.Introduction

Metamaterial (MM) is an arrangement of artificial structure elements. It can achieve advantageous and unusual electromagnetic properties, which cannot be realized with natural material [1]. These properties provide new alternatives to manipulate the propagation characteristics of electromagnetic waves, and give rise to a variety of applications, such as planar superlens and invisible cloaks in microwave as well as the THz functional devices [2-6].

Metallic structures on dielectric substrates are commonly used in MMs [7]. They have relative low losses at microwaves, and allow us to demonstrate experimentally the extraordinary properties of the MMs. However, as the frequency is pushed higher towards the terahertz (THz), the ohmic losses become prominent and current MM design may not implement the desired functions. Thus, an urgent problem is to reduce the loss of THz MMs. Recently, the metallic THz MM operating at cryogenic temperature and superconducting THz MMs are proposed for this purpose [8-9]. The simulations show the loss of the MM can be reduced by 40% as the normal metal is at 1K. For the superconducting THz MM made of YBCO film, the experiments show the loss decreases as the temperature decreases.

However, the low loss behavior is not displayed completely because measuring system can not cool down the sample below the superconducting transition temperature $T_c$. Moreover, YBCO film has a comparable surface resistance $R_s$ to the normal metal at 100 GHz and 77 K, which implies that YBCO film may not a good candidate for superconducting THz MMs [10].

The first topic in this Letter is to study THz MM made of superconducting Nb films. It has been reported that the $R_s$ value of Nb film at 4.2 K is at least one order lower than that of YBCO film up to 300 GHz [10]. This means the loss of Nb THz MM can be greatly reduced. Indeed, our measurement shows the quality factor can be about three times as large as the one made of normal metal at 1 K [8].

Besides the low loss, magnetic-field tuning is a unique property for superconducting MMs. In the earlier days, this tuning possibility was demonstrated at microwave frequencies [11-12]. For the potential applications, it would be highly desirable to push this tuning effect up to THz and provide another alternative to control THz wave propagation apart from electric[2], optical[13-14] and thermal tuning [15]. The second topic in this Letter is to demonstrate the magnetic field tuning on superconducting THz MM.

## 2. Experiments and discussions

A square split ring resonators (SRRs) array, as shown in the inset of Fig.1 is proposed for our study [7,16]. A double-ring SRR consists of two concentric rings each interrupted by a small gap. The unit cell is of the length of 140 μm. The side length of outer ring is 120 μm. The width of the ring and distance between two rings are all 10 μm. The gap distance is also 10 μm. The transmission property of this MM was simulated by the commercially available code. In the simulation, the superconducting thin film is assumed to be a normal metal. The wave propagates perpendicular to the plane of SRR, with the ac electric field parallel to the gap (shown in Fig.1). Three absorption dips below 0.5 THz are observed which resonate at 132, 240 and 420 GHz, respectively.

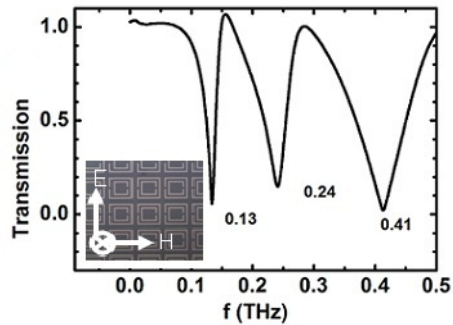

Fig.1 The simulated transmission spectrum of our THz metamaterial. The microscopic

image of the sample array is in the inset.

Superconducting Nb films are deposited on 400 μm Si (111) substrate by DC magnetron sputtering. $T_c$ is measured by four-probe technique, giving a value of about 9 K. The crystal structure of Nb films was studied by XRD. The films are polycrystalline with a cubic structure, and the main diffraction pattern corresponding to Nb (110) planes parallel to the substrate surface. The standard photolithography and plasma etching are used to form the pattern.

Transmittance experiments were carried out by a continues-wave THz spectroscopy with a superconducting split-coil magnet [17]. We measured transmission spectra at 6 K and 26 K in two frequency ranges, one from 100 to 180 GHz and the other from 300 to 550 GHz, which corresponds to the first and third resonant modes. In the case of zero magnetic fields and at T = 6 K we observed respectively two sharp absorptions at about 132 GHz (Fig. 2) and 418 GHz (Fig. 3), which agree with the simulations shown in Fig.1. The resonance curves can be expressed as [18]

$$P_r(f)/P_r(f_0) = \frac{1+(1+\beta/2)^2 Q_L^2 (f/f_0 - f_0/f)^2}{1+Q_L^2 (f/f_0 - f_0/f)^2},$$

where $P_r(f)$ is the received power, $f_0$ is the resonant frequency, $Q_L$ is the loaded quality factor and $\beta$ is the coupling coefficient. Unloaded quality factors ($Q_u$) can be deduced by $Q_u=(1+\beta/2) Q_L$. Using Eq. (1) to fit the measured results as shown in Fig.2 and 3 (solid lines), we get $Q_L=5$ and $1+\beta/2=17.3$, yielding $Q_u=86.5$ for resonating at 132 GHz, and $Q_L=3$ and $1+\beta/2=12.3$, yielding $Q_u=36.9$ for resonating at 418 GHz, respectively. Recent simulation shows that the $Q_u$ values of THz MM made of alumina (Al) at 1 K is 10.5 at 650 GHz [8]. Since the surface resistance is proportional to the square root of frequency for normal metal, the $Q_u$ values of the Al film MM is estimated about 23 at 132 GHz and 13 at 418 GHz. Obviously, the $Q_u$ values of our THz MMs is about three times as large as the one made of normal metal. The physical reason is clear because superconducting Nb film has a lower surface resistance than the normal metal. At 6 K, the complex conductivities of Nb film from the previous measurement are about $(2-j6)\times 10^5$ $\Omega^{-1}$ cm$^{-1}$ at about 150 GHz (close to 132 GHz) and $(1.2-j3)\times 10^5$ $\Omega^{-1}$ cm$^{-1}$[17]. From these values, $R_s$ are about 2.6 mΩ at 150 GHz and 7.9 mΩ at 450 GHz. For bulk Al, the conductivity is 0.4 GS/m at 1 K, and the calculated Rs are 4.5 mΩ at 132 GHz and 8.4 mΩ at 450 GHz. Our results demonstrate clearly that the superconducting THz MM is a very nice candidate to achieve low loss performance.

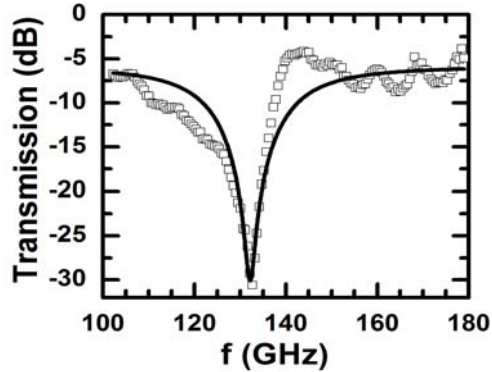

Fig.2  Transmission spectrum at 6 K and zero magnetic field (Hdc) for resonance mode at 132 GHz. The solid line represents the fit to the experimental results (symbols)

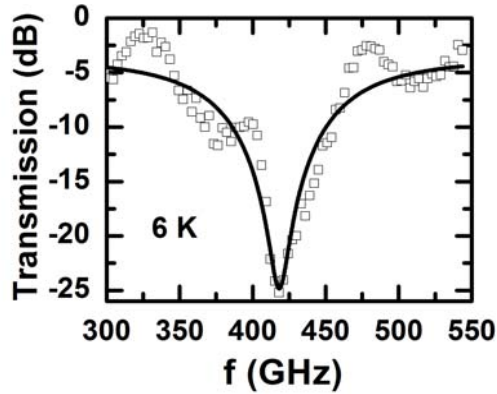

Fig.3 Transmission spectrum at 6 K and zero Hdc for resonance mode at 418 GHz. The solid line represents the fit to the experimental results (symbols)

In addition to the low loss properties, superconducting MM can implement DC magnetic field tuning. Fig. 4 and 5 showed the transmission spectra at 6 K and different magnetic field (0-1 Tesla) for two resonance modes. Obviously, the resonant frequencies change as $H_{dc}$ increases from zero to 0.7 T. After that, the resonant frequencies keep constant. The physical reason of tuning is that the superconducting properties of Nb film, such as the magnetic penetration depth and critical current density strongly depend upon $H_{dc}$ [12,19]. As $H_{dc}$ is larger than the upper critical field, $H_{c2}$, the superconductivity of Nb film is destroyed, and the film goes into the normal state and the film properties do not depend on $H_{dc}$. To verify the described effects the transmission spectra at 26 K were also measured (symbols). We found that these spectra are almost the same as the spectra at 6 K and 0.7 T. These measurements mean that $H_{c2}$ of our film is about 0.7 T. Reported values of the critical field at low temperatures range from 1.0 T to 4.6 T [20]. According to the temperature dependence of $H_{c2}$, i.e., $H_{c2}(T)= H_{c2}(0)(1-t^2)/(1+t^2)$ [19], where $t=T/T_c$ and T is working temperature, we can obtain $H_{c2}(6\ K)$ is in a range from 0.4 T to 1.8 T, which covers our estimation.

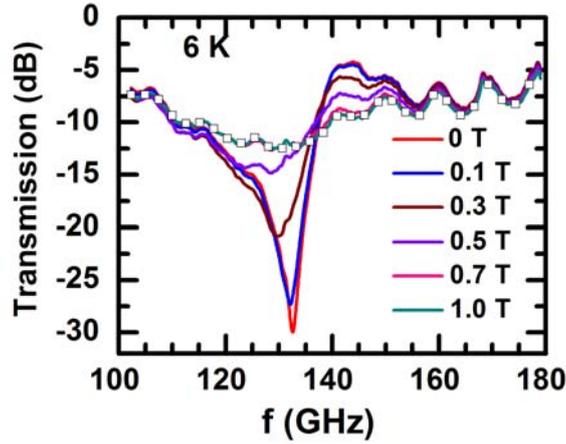

Fig.4 The solid lines represent theransmission spectra at 6 K and $H_{dc}$=0, 0.1, 0.3, 0.5, 0.7 and 1 T (started from bottom), respectively. The symbols represent the transmission at 26 K and zero $H_{dc}$.

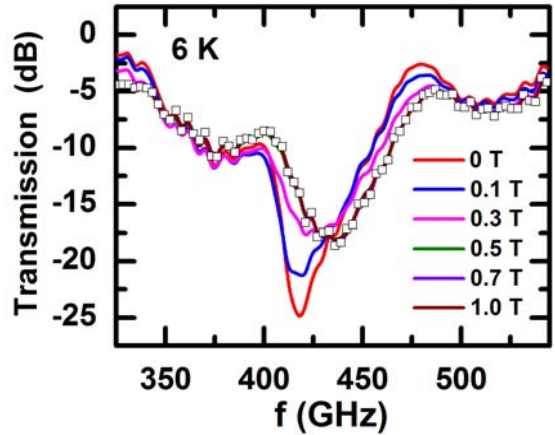

Fig.5 The solid lines represent theransmission spectra at 6 K and $H_{dc}$=0, 0.1, 0.3, 0.5, 0.7 and 1 T (started from bottom), respectively. The symbols represent the transmission at 26 K and zero $H_{dc}$.

We also observed different frequency-tuning directions for two resonant modes. For the resonant mode at 132 GHz, the frequency decreases with the increasing $H_{dc}$. The total frequency change is about 3 GHz. But, for the resonant mode at 420 GHz, the frequency increases with $H_{dc}$ with a total frequency change of about 20 GHz. The different tuning directions come from the different current distributions of the different resonant modes. Fig.6 shows the current distributions from electromagnetic field simulation. For the resonant mode at 132 GHz, there is a circulating current in the outer ring. In this case, this ring acts as an inductor and the gap is like a capacitor [20]. The magnetic penetration depth of superconducting film increases with $H_{dc}$, leading to the increase of the inductor, and decrease of the frequency. For the absorption at 418 GHz, the current distribution shows symmetric style. The current in the outer ring has a same direction with the one in the adjacent our ring, leading to attractive forces between them. As the film goes to the superconducting state, this attraction become larger, and the current prefers to distribute in the outer edge of the outer ring. Effectively, the electric length of the outer ring is extended, and the resonant frequency is reduced.

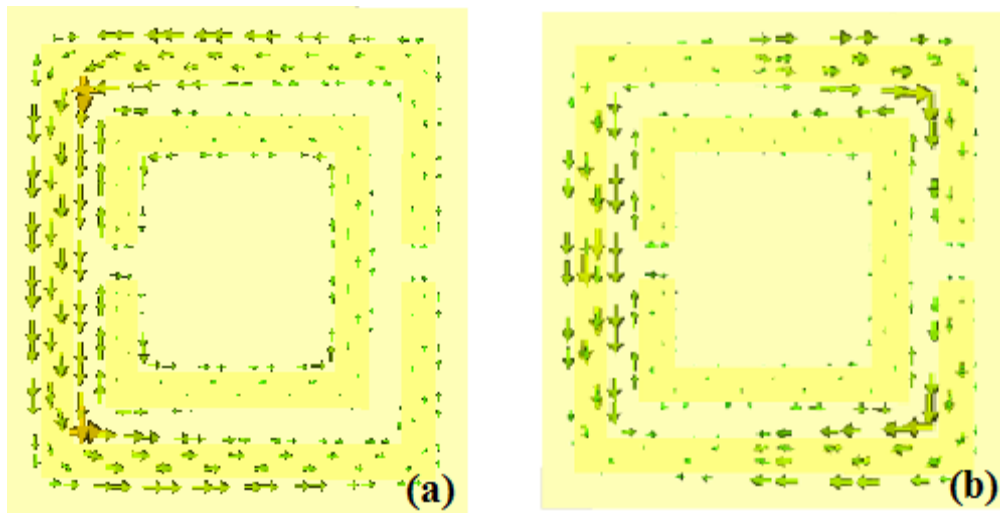

Fig.6. The surface current distribution for resonant modes (a) at 132 GHz and (b) at 420 GHz

## 3. Summary

In conclusion, we successfully demonstrated the low loss and magnetic field tuning superconducting THz MM. The Q values of the MM at 132 GHz and 418 GHz is about three times as large as the one made of Al film at 1K, which is due to the low surface resistance of superconducting Nb film. We hope our results can offer new path to make THz devices based on superconducting THz MM.

**Acknowledgments**

This work is supported by the MOST 973 project (No.2007CB310404) of China, the Program for New Century Excellent Talents in University (NCET-07-0414), Ministry of Education, China, and Qing Lan Program of JiangSu Province, China.